\documentclass{rsproca}%%%%where rsproca is the template name

\usepackage{graphicx,subfigure,amsmath,bbm}
\usepackage{tabularx}
\usepackage[square,comma,sort&compress,numbers]{natbib}
\usepackage{epsfig}
\usepackage{graphicx}% Include figure files
\usepackage{dcolumn}% Align table columns on decimal point
\usepackage{bm}% bold math
\usepackage{amssymb}
\jname{rsif}
\Journal{J. R. Soc. Interface\ }

\begin{document}

\title{Modular knowledge systems accelerate human migration in asymmetric random
environments}

\author{%%%% Author details
Dong Wang$^{1}$ and Michael W. Deem$^{1,2,3}$}

%%%%%%%%% Insert author address here
\address{$^{1}$Department of Physics \& Astronomy, Rice University,
Houston, TX 77005, USA\\
$^{2}$Department of Bioengineering, Rice University, Houston, TX 77005, USA\\
$^{3}$Center for Theoretical Biological Physics, Rice University, Houston, TX 77005, USA}
\keywords{modularity, human migration, asymmetry, Americas}
\corres{Michael W. Deem\\
\email{mwdeem@rice.edu}}

\begin{abstract}

Migration is a key mechanism for expansion of communities.  
In spatially heterogeneous environments, rapidly gaining knowledge about
the local environment is key to the evolutionary
 success of a migrating population. For historical human migration, 
environmental heterogeneity was naturally asymmetric in the north-south (NS)
 and east-west (EW) directions. We here consider 
the human migration process in the Americas, modeled as random, 
asymmetric, modularly correlated environments. Knowledge about
 the environments determines the fitness of each individual.  
We present a phase diagram for asymmetry of migration as a function
of carrying capacity and fitness threshold.
We find that the speed of migration is proportional
 to the inverse complement of the spatial environmental gradient, and in particular
 we find that north-south migration rates are lower than 
east-west migration rates when the environmental gradient is higher in the 
north-south direction.  Communication 
of knowledge between individuals can help to
 spread beneficial knowledge within the population. The speed of migration increases 
when communication transmits pieces of knowledge that contribute in a modular way
 to the fitness of individuals.
The results for the dependence of migration rate on asymmetry
 and modularity are consistent with existing archaeological observations. 
The results for asymmetry of genetic divergence are consistent with patterns 
of human gene flow.

\end{abstract}

\maketitle

\section{Introduction}

Interesting phenomena emerge in the population dynamics
 in heterogeneous environments. For example, experimental and theoretical
 studies have shown that spatial heterogeneity accelerates
 the emergence of drug resistance 
\cite{PhysRevLett.105.248104,zhang2011acceleration} and solid tumor evolution in heterogeneous  microenvironments
 \cite{zhang2012physics}. On a larger scale, heterogeneity
 plays a central role in population biology of infectious 
diseases \cite{PhysRevLett.99.148701} 
and emerges in the development
 of large physics projects, such as ATLAS, CERN \cite{Turtscher}. Finally, in heterogeneous environments,
evolved networks are modular when there are local extinctions
 \cite{kashtan2009extinctions}.

Populations experience heterogeneous environments during
 migration.  Migration can occur in different dimensions:
 for example, cells undergo one-dimensional, 
two-dimensional, or three-dimensional migration \cite{doyle2009one}.
 In two-dimensional or three-dimensional migration, 
the environmental gradient can additionally
be distinct in the different directions.
 For example, in the case of human migration, 
the north-south direction has a greater environmental 
gradient than does the east-west direction \cite{diamond1997guns}.
The heterogeneity is important in simulating human dispersal in 
the Americas \cite{steele2009human}.
 In the east-west direction, food production spread from southwest Asia 
to Egypt and Europe at about $0.7$ miles per year around 5000 BC, while in 
the north-south direction, it spread 
northward in the American continent at about $0.2$ to $0.5$
 miles per year around 2000 BC \cite{diamond1997guns}. This spread is on the same
 order as the velocity of human migration, so we estimate 
that the human migration velocity in the east-west direction is
 about $2$ to $3$ times faster than in the north-south direction.
Previous work has generated detailed migration paths using geographical 
data \cite{anderson2000paleoindian} as well as results that match existing archaeological evidences 
well after considering spatial and temporal variations \cite{steele2009human}. 
We do not try to generate a detailed map of human migration in this paper. 
Instead, we use a general model to generate east-west north-south asymmetry 
and study the role of a modular knowledge system.

Knowledge of local environments, such as 
effective agricultural or animal husbandry techniques,
was vital to the survival of these early migrants \cite{diamond1997guns}.
Evolutionary epistemology views the 
gaining of knowledge as an adaptive process with blind 
variation and selective retention \cite{campbell1960blind}. 
Communication of
 knowledge between individuals is also 
an efficient means to spread this discovered, locally
adapted knowledge \cite{mithen1996preshistory}. Similarly, models of social 
learning theory stress the importance of social learning in the spread of 
innovations \cite{kandler2010social}. Here we
model the adaptation of a population to the local environment
 using an evolutionary model with natural selection, mutation and communication. 
The knowledge of an individual
 determines his or her fitness. Evolutionary psychology 
and archeology posit that the human mind is modular \cite{steele1996weak},
 and that this modularity is shaped by evolution 
\cite{tooby1995psychological} and facilitates  understanding of local environments
\cite{mithen1996preshistory}. Conjugate to this modularity
 must be dynamical exchange of corpora of knowledge between individuals \cite{Goldenfeld2011,AR}. 

\section{Methods}

\begin{table}[h]
\caption{Symbols used in this paper}\label{Table1}
\begin{tabular}{|c|c|}
\hline
Symbol & Meaning \\\hline
$\chi$ & Similarity between adjacent environments \\\hline
 $v$ &Emigration velocity \\\hline
 $t$ & Emigration time\\ \hline
$N$ &Number of individuals in one environment\\\hline
$N^*$ &Carrying capacity of one environment\\\hline
$N_0$ &Initial population size of one environment\\\hline
$f$ & Fitness\\\hline
$f^*$ &Fitness threshold\\\hline
$J$ & Interaction matrix\\\hline
$\Delta$ & Connection matrix\\\hline
 $K$ & Number of modules in a sequence\\\hline
$l$ & Module size \\\hline
$\mu$ & Mutational rate\\\hline
$\nu$ & Knowledge transfer rate\\\hline
$d$ &Genetic distance\\\hline
 $S$ & A whole sequence\\\hline
$s$ & One locus in a sequence \\\hline
$L$ & Length of one sequence\\\hline
$M$ & Modularity\\\hline
\end{tabular}
\end{table}

\begin{figure}
\begin{center}
\includegraphics[scale=0.85]{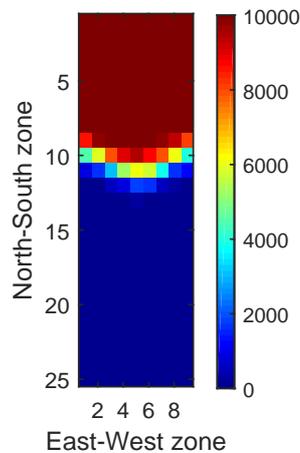}
\end{center}
\caption{\label{Fig1}
Population distribution half way through 
the migration. Color
 indicates density of the population in each environment. The maximum 
capacity of each environments is $N^*=10000$. Initially there are 
$1000$ individuals in the top center environment  $(1,5)$, and no individuals in
 other environments. Here $\chi_{\text{EW}}=0.8$, 
$\chi_{\text{NS}}=0.4$, $f^*-2L=0.3 L$, $L=100$, $\mu=0.5$, 
$\nu=6$, and $K=5$. Density was averaged over $24$ runs.
}
\end{figure}

Table \ref{Table1} shows the symbols in this paper. 
The observed emigration time and asymmetry of emigration 
time are critical in the determination of the values of 
these parameters.
We consider migration in random, asymmetric, 
modularly correlated environments.  
We use $9 \times 25$ correlated, random environments, 
where $25$ is the number of environments
 in the north-south direction at the same longitude \cite{kottek2006world}, 
and $9$ is chosen so that 
$9/25$ is approximately the ratio of the east-west to 
north-south dimension of the
 Americas. See Fig.\ \ref{Fig1} for an illustration,
 where each square block corresponds to an environment.

Each individual 
$\textbf{a}$ has a fitness $f_\textbf{a}$, as well as a sequence
 $S^\textbf{a}$ that is composed of $L$
loci, $s_i^\textbf{a}$, representing the knowledge of the 
individual. Fitness describes reproductive success and is proportional to
the reproduction rate. For simplicity, we take $s_i^\textbf{a} = \pm 1$.
We first consider a linear fitness landscape, later
generalizing to an interacting landscape:
\begin{eqnarray}
f[S] &=& 2 L + H[S]
\nonumber \\
H[S] &=& \sum_{i} s_i J_{i} 
\label{Eq2}
\end{eqnarray}
where $J_{i}$ is a quenched, Gaussian random interaction parameter, 
with variance $\sqrt 2$, and the offset $2L$ is chosen
so that fitness is non-negative, since $H_{\min}$ is $-2 L /\sqrt\pi$. 
For a given instance of the model, the interaction parameters $J_{i}$ 
are randomly chosen and then fixed for that instance of the model.
When for each $i$ from $1$ to $L$, $s_i J_i>0$, the fitness reaches 
its highest value, and natural selection selects the sequence with 
the best configuration.

The fitness of the population is influenced by the environment, 
quantified by interaction parameters $J$, describing the interaction between 
 knowledge element $i$ of the individuals and the environment
(see also Eq.\ \ref{Eq2} above). The 
interaction parameters $J$
in two adjacent environments, $J$ and $J'$, are correlated,
\begin{equation}\label{Eq1}
\langle J_{i} J'_{i}\rangle/\langle J_i^2\rangle = \chi
\end{equation}
where $\chi = \chi_{\rm{EW}}$ if the two have the same latitude,
 and $\chi = \chi_{\rm{NS}}$ if they have the same longitude. The
 smaller the $\chi$, the bigger the environmental gradient is. Here $0<\chi<1$,
and $\chi_{\rm{NS}} < \chi_{\rm{EW}}$, since the gradient of environment in the
north-south direction is more dramatic \cite{diamond1997guns}. 

In each environment, we use a Markov process to describe the
 evolutionary dynamics, including replication with rate $f$, mutation
with rate $\mu$ coming from discovering new knowledge
 through trial and error, and transfer of a corpus of knowledge
of length $L/K$ with rate $\nu$. When individuals reproduce, 
they inherent the knowledge and genes from their parent without error.
Both mutations and knowledge transfers are random, and they do not
depend on the fitness of individuals.
 The relative rates of replication, mutation, and transfer
 are $f$, $\mu L$, and $\nu K$, respectively, so on average each individual
 makes $\mu L/f \approx \mu/2$ mutations, 
as $f \approx 2L$ at short times for which these populations evolve,
and $\nu K/f \approx \nu K/(2L)$ 
knowledge transfers per lifetime of an individual.
We set the information sequence length $L=100$.
Discovery of new facts, represented by mutation, changes
one site, or 1\% of the knowledge of an individual,
whereas knowledge transfer changes $1/K$ of the knowledge.
Discovery of new facts should be rare,
and in our simulation 
we set $\mu=0.5$, so that approximately one-quarter of the 
individuals attempt to make a 
discovery through trial and error during his or her 
lifetime.
We consider $K=5$ corpora 
of knowledge. Transfer of one corpus, for example, could be one farmer
attempting to communicate to another farmer how to grow a new crop in a new environment. 
Knowledge transfer must be
rare, so we set $\nu=6$, so that roughly $\nu K / (2 L) \approx 1/7$ 
of the individuals attempt a
knowledge transfer process during his or her lifetime. 
We additionally consider various values of $\nu$ in 
this work to investigate the coupling of $\nu$ to modularity.
Selection is based on the fitness of the knowledge and
it determines the the utility of theses mutation and knowledge transfer
events.

This dynamics of migration is described by a
Markov process, whose master equation
is detailed in the Appendix.
Initially, one of the 
environments with the highest latitude is occupied by $1000$ 
individuals with random sequences, as Native Americans are believed
 to have entered the Americas through Alaska in the north. Since
 the population migrates from north down to south, we only allow 
migration to the east, west, and south. In each environment, the 
population evolves according to the Markov dynamics.
% described by equation \ref{Eqn.2}. 

The qualitative behavior of the migration depends on the
carrying capacity, $N^*$, and the fitness threshold, $f^*$. The carrying capacity is defined 
as the maximum population load of an environment \cite{hui2006carrying}. 
After the population size reaches $N^*$, we randomly kill an individual every time another 
individual reproduces, as described in detail in Eq.\ \ref{Eq8}. As a result, the total 
number of individuals does not exceed $N^*$.
The initial colonization of the Americas 
occurred before the Common Era, for which there are no reliable population data. 
\begin{figure}
\begin{center}
\includegraphics[scale=0.7]{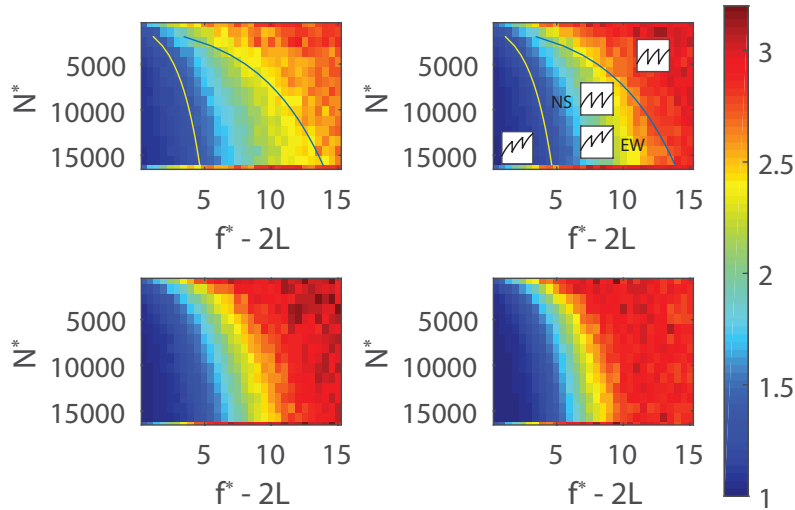}
\end{center}
\caption{\label{Fig2}
Asymmetry in emigration times $t_{\text{NS}}/t_{\text{EW}}$ for different $N^*$
 and $f^*$.
Upper left, phase diagram of the linear model, Eq.\ \ref{Eq2}.
Upper right, phase diagram of the quadratic model, Eq.\ \ref{Eq5}, with $M=1$.
Lower left, phase diagram of the quadratic model, with $M=1/2$.
Lower right, phase diagram of the quadratic model, with $M=0$.
Other parameters are as in Fig.\ \ref{Fig1}.
The color indicates the asymmetry in emigration times:
$t_{\rm NS} / t_{\rm EW}$. 
There are three phases, with the boundaries denoted by the two curves.
The steady-state fitness dynamics, $f(t)$ vs $t$, of the right 
phase and the left phase are shown in inset. The fitness dynamics of the middle
phase in the north-south direction follows that of the upper inset, 
and in the east-west direction
 follows that of the lower inset.  The phase boundaries are given approximately by
equating the times in Eq.\ \ref{Eq3} for the north-south (left) or 
east-west (right) migration directions for linear model and $M=1$ quadratic model.
The model for human migration has $N^* = 10000$ and $f^*-2L = 30$.
}
\end{figure}

It is estimated that there were seven million people in the Americas at the start 
of the Common Era \cite{maddison2007world}, corresponding to $7000000/(25\times 9)=31111$ 
individuals in each environment. We choose the carrying capacity to be $N^*=10000$, less
than $31111$, reflecting that the population size was smaller the earlier time of initial 
population expansion. 
We show the results for various $N^*$ in Fig.\ \ref{Fig2}. 
We introduce the fitness threshold, $f^*$, 
because individuals need to be well prepared before emigrating to the next environment. 
For example, young male ground squirrels appear to disperse after attaining a threshold 
body mass \cite{nunes1996mass}, and dispersing males tend to have greater fat percentage for 
their bodies \cite{nunes1996mass}. The increased body mass and fat 
percentage are thresholds required for migration. Similarly, naked mole-rats migrate more 
frequently after body mass reaches a certain value \cite{o1996dispersive}. 
It is possible that some individuals try to emigrate without reaching the fitness 
threshold when the local population size reach environmental capacity. 
However, they are not fit enough to colonize the new environment. 
Thus, we employ a fitness threshold in our approach, and allow no emigration 
before the average fitness value reaches $f^*$. 
When the population size reaches $N^*$ and 
the average fitness reaches $f^*$ in an environment, we 
move $N_0=1000$ randomly chosen individuals to one of the unoccupied adjacent
 environments. Fitter individuals
may be more likely to migrate since they are physically better prepared to
migrate, while on the other hand less fit individuals may have more desire
to migrate since they do not live well in the current environment. We randomly 
choose individuals to migrate because of this ambiguous relationship between 
fitness and migration.
If we move fitter individuals instead of 
randomly chosen individuals, the
initial fitness of the individuals in the new environment will be higher.
Thus, effectively the $\chi$ would be higher. 
The time required for a population to emigrate from
 an environment is denoted by the emigration time, $t$, 
 and the emigration velocity $v$ is defined as $v=1/t$. The emigration time 
of an environment is the time from the arrival of the first individuals to 
the departure of the first individuals.

To compare our results with current human genetic data, we assign to each 
individual another sequence $S'$, also composed of $L$ loci, and each locus
can take values $\pm 1$. These sites correspond to automosal microsatellite 
marker genotype data \cite{wang2007genetic}, which we will compare with later in this paper. 
The traits of the genetic data are neutral in 
our model. That is, the values of the loci in the sequence $S'$ have no 
effect on the fitness. The genetic sequence mutates at a rate $\mu'$. 
When an individual reproduces, both the knowledge sequence and the 
genetic sequence reproduce.

\begin{figure}
\begin{center}
\includegraphics[scale=0.5]{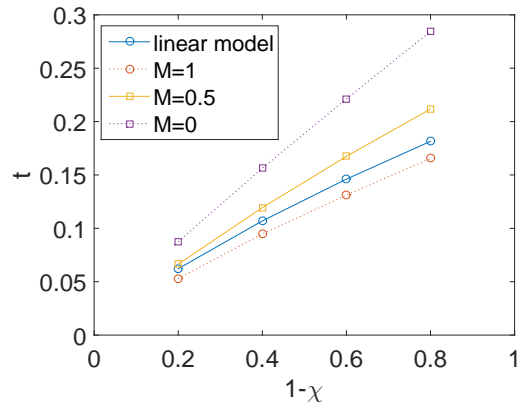}
\end{center}
\caption{\label{Fig3} Emigration time versus $1-\chi$ 
 for the linear model and the quadratic model with various modularities.
 Here $\chi_{\rm EW} = \chi_{\rm NS} = \chi$, and other
 parameters are as in Fig.\ \ref{Fig1}. }
\end{figure}

We set the time scale in our simulation by the observation that
 Native Americans spent about $10000$ years to migrate from the north
 tip of the American continent to the south tip \cite{goebel2008late}, experiencing 
about $25$ climate zones \cite{kottek2006world}, so migration
 to a new environment occurred roughly every $400$ years, i.e. roughly every $20$ 
generations. In our simulation, we define a generation as the time
 period during which on average, each individual is replaced by 
another individual. We find that the population migrates approximately
 once per $20$ generations when $\chi_{\text{EW}}=0.8$, 
$\chi_{\text{NS}}=0.4$, and $f^*-2L=0.3L$. One can estimate 
how many generations it takes to migrate to the next environment. The rate of 
change of fitness at short time roughly follows \cite{park2015modularity},
\begin{equation}
df/dt = 2L
\end{equation}
Since $\Delta f=0.6\times(f^*-2L)=0.18\times 2L$ for migration from the north 
or 
$\Delta f=0.2\times(f^*-2L)=0.06\times 2L$ for migrating 
from the east or west, the emigration time is $0.18$ or $0.06$ 
depending on the origin of migration, and this is consistent with Fig.\ \ref{Fig3}.
We use $\Delta t=0.1$ as a rough estimate for emigration time.
To convert this time in our simulation to number of human replications, we consider 
that one replication takes around $dt=1/f=1/2L$ time, so one emigration takes 
$\Delta t/dt=20$ generations.
To compare the genetic data with 
current human data, we allow all 
environments to evolve for another $10000$ 
years after all environments are occupied, without migration between environments. 
We assume no gene flows between these environments, as previous work \cite{ramachandran2011test}
assumes that the asymmetry
in the genetic distance originates from the asymmetry of gene flows in
different directions. Here we investigate another possible origin of the
asymmetry of genetic distance, that is, the asymmetry already exists
when the population colonized the Americas. It is quite possible
that both mechanisms help to create this asymmetry, but in order to show that the
initial colonizing process itself could generate this asymmetry, we 
suppress the possibly asymmetric genetic flows.

\section{Results}\label{sec:results}

In Fig.\ \ref{Fig1} we show a snapshot of population 
distribution, approximately half way through the migration. 
Migration sweeps south
 and spreads both to the east and west. Migration forms a tilted 
front, with slope magnitude equal to $v_{\text{NS}}/v_{\text{EW}} = 0.35$,
indicating the velocities of migration in 
different directions are different. 

In Fig.\ \ref{Fig2}
we show the three possible phases for different carrying capacity, $N^*$, and 
fitness threshold, $f^*$. Different phases correspond to whether 
the migration is limited by the fitness
threshold or the population size threshold. In the left phase, the
population is limited by the population size threshold, and there is
no east-west north-south asymmetry. In addition, as the population
migrates, the maximum fitness value increases since the population is
allowed to evolve further after reaching the fitness threshold, as shown
in the left inset of the upper right figure. In the middle phase, the
migration in the east-west direction is limited by population size
threshold while the migration in the north-south direction is limited
by the fitness threshold. The maximum fitness value increase as the
population migrates in the east-west direction, but in the north-south
direction, the maximum fitness value is $f^*$. The degree of the
east-west north-south asymmetry increases in this phase from the boundary
with the left phase to the boundary with the right phase. In the right
phase, migrations in both directions are limited by the fitness
threshold, and the maximum fitness value remains the same as the
population migrates. The east-west north-south asymmetry is approximately
unchanged in this phase.
The boundaries of these phases are determined by
noting the times to reach the carrying capacity
and the fitness threshold:
\begin{eqnarray}
t_{N^*} &=& \frac{\ln (N^* / N_0) }{2 L}
\nonumber \\
t_f &=& \frac{(f^*-2L) (1-\chi) }{2 L}
\label{Eq3}
\end{eqnarray}
where $N_0$ is the initial population of one environment. 
Here we have used that the evolution  of the
fitness in one generation is small compared to the
offset $2 L$, and that the evolution within one
environment at steady state is from $\chi (f^*-2L)+2L$
to $f^*$ in the rightmost phase.
The left phase boundary in Fig.\ \ref{Fig2} is
given by the condition $t_{N^*}  = t_f$ in the north-south direction, and
the right phase boundary is given by $t_{N^*}  = t_f$ in the east-west direction. 
We note that our current choice of parameters is deep in the right phase, 
indicating that the east-west north-south asymmetry is robust to the change of 
$f^*$ or the ratio $N^*/N_0$.

We determine quantitatively how the environmental gradient 
influences the velocity of migration. In Fig.\ \ref{Fig3} we show the
 emigration time versus $1-\chi$, the change between adjacent 
environments.
 It is interesting that the emigration time is approximately 
proportional to $1-\chi$. This occurs because in our simulation 
for these parameters,
 the population reaches $N^*$ earlier than $f^*$, so the emigration
 time is the time required to reach $f^*$. 
For our model,
 $f^*-2L = 0.3L$, while $\max(f-2L)\approx 2L$, 
so $f^*$ is still far from optimal, and the
 fitness increases linearly with time in the regime we are discussing.
 So $t=\Delta f / v_f = (1-\chi) f^* /v_f$, where $v_f$ is a 
constant for a fixed modularity. So $t\propto 1-\chi$, and we 
quantify the ratio of velocity in the two different directions as
\begin{equation}\label{Eq4}
\frac{v_{\text{EW}}(M)}{ v_{\text{NS}}(M)}=\frac{1/t_{\text{EW}}(M)}
{1/t_{\text{NS}}(M)}=\frac{1-\chi_{\text{NS}}}{1-\chi_{\text{EW}}}
\end{equation}

In the linear model, it is quite easy to evolve the optimal
pieces of knowledge, while in reality, finding the best knowledge
is difficult at the individual level. 
We now show that these results are robust to considering an
interacting model, while also demonstrating the significance
of the modularity order parameter in the interacting model.
As finding optimal knowledge
 for a local environment is difficult, the fitness 
landscape is rugged \cite{PhysRevLett.99.228107}, and we use
a spin glass to represent the fitness:
\begin{eqnarray}\label{Eq5}
f[S] &=& 2 L + H[S]
\nonumber \\
H[S] &=& \sum_{ij} s_i s_j J_{ij} \Delta_{ij}
\end{eqnarray}
where $J_{ij}$ is a Gaussian random matrix, 
with variance $1/C$.
The offset value $2 L$ is chosen by Wigner's semicircle law
 \cite{wigner1958distribution} so that the
minimum eigenvalue of $f$ is non-negative. 
The entries in the matrix $\Delta$ are zero or one, with probability
$C/L$ per entry, so that the average number of connections
 per row is $C$. The optimization of this fitness model is hard
when $L$ is large, and here we give a simple example to show why. 
Consider a case when $J_{ij}>0$, $J_{ik} >0$ and $J_{jk}<0$ 
given $i$, $j$ and $k$. To make $J_{ij}s_is_j$ positive and fitness value 
larger, $s_i$ and $s_j$ must have the same sign. Similarly, to make $J_{ik}s_is_k$ 
and $J_{jk}s_js_k$ positive, we need $s_k$ to have the same sign with $s_i$, 
and $s_k$ to have different sign with $s_j$. This indicates that $s_i$ and $s_j$ 
have different signs, contradicting that $s_i$ and $s_j$ having the same sign. This 
phenomena is called frustration in physics \cite{sadoc2006geometrical}, making the  
fitness hard to optimize. Let us exemplify this using an example from the human knowledge 
system. Humans developed three pieces of knowledge: the knowledge of toxicity of 
mushrooms, the knowledge of red food, and the knowledge of apples. The interaction 
between the first two pieces of knowledge implies that red food is bad and 
undesirable, while the later two pieces of knowledge implies something on the 
contrary. As a result the human knowledge system can be difficult to optimize.  
We will discuss how modularity helps to reduce this frustration and thus 
makes it easier to optimize the fitness in section \ref{sec:discussion}.

\begin{figure}
\begin{center}
\includegraphics[scale=0.7]{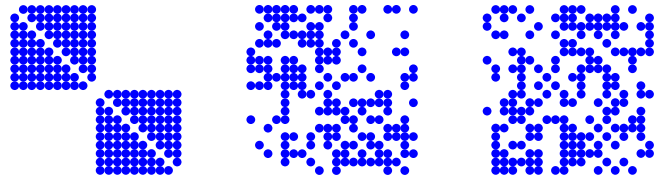}
\end{center}
\caption{\label{Fig6} Illustration of $L=20$ connection matrices with 
different modularities. Left, a completely modular connection matrix, $M=1$. Middle, 
a moderately modular connection matrix, $M=0.5$. Right, a non-modular 
connection matrix, $M=0$.}
\end{figure}

We introduce modularity by an excess of interactions in
 $\Delta$ along the
$l \times l$ block diagonals of the $L \times L$
connection matrix. There are $K$ of these
block diagonals, and $K=L/l$.  Thus,
the probability of a connection is
$C_0/L$ when
$ \lfloor i/l \rfloor \ne \lfloor j/l \rfloor$
and $C_1/L$ when
$ \lfloor i/l \rfloor = \lfloor  j/l \rfloor$.  The number of
connections is $C = C_0   + (C_1 - C_0) /K$, and modularity 
is defined by $M = (C_1 - C_0)  / (K C)$. In Fig.\ \ref{Fig6} we illustrate 
three $20\times 20$ matrices with modularities $1$, $0.5$ and $0$ and $C=9$.

\begin{figure}
\begin{center}
\includegraphics[scale=0.62]{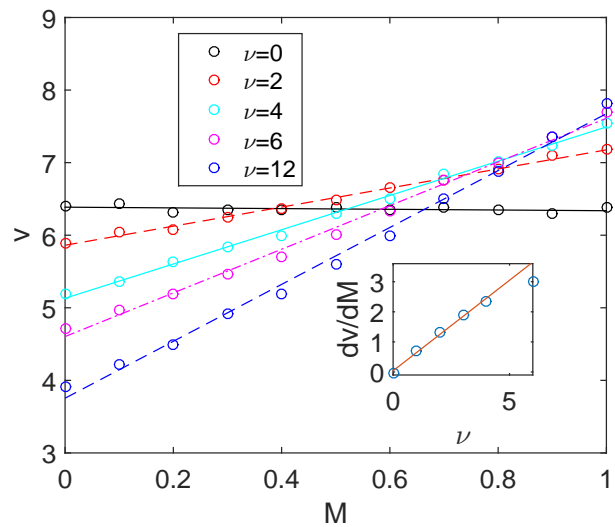}
\end{center}
\caption{\label{Fig4} North-south Emigration velocity versus modularity for 
different $\nu$. The lines are linear fit of the data of the corresponding 
horizontal gene transfer rate.
The inset shows $dv_{\rm NS}/dM$ versus $\nu$.
 The dots are data points and the line is a linear fit to the data. 
Other parameters are as in Fig.\ \ref{Fig1}. }
\end{figure}

Modularity, coupled with knowledge transfer, accelerates the 
evolution of a population in a new environment \cite{park2015modularity}.
 We now check how modularity and knowledge transfer influence the
 velocity of migration. For different $M$ and
 $\nu$, the results are shown in Fig.\ \ref{Fig4}. For small $M$, a larger
 $\nu$ implies a smaller migrating velocity, indicating that the transfer of
(non-useful) knowledge slows down evolution. As modularity increases,
 the migration velocity at larger $\nu$ catches up with that of smaller
 $\nu$. At $M=1$, in the range of $\nu$ shown, 
the faster the population migrates faster for larger $\nu$. 

We fit the curve of $v_{\rm NS}$-$M$ for $\nu\leqslant 4$ in Fig.\ \ref{Fig4} with 
linear regression, observing $R^2\geqslant 0.95$, except for
 $\nu =0$, which has a zero slope and larger noise. We also fit the data
 for $\nu=1$ and $\nu=3$, not shown in Fig.\ \ref{Fig4}. We show 
$dv_{\text{NS}}/dM$ versus modularity for different $\nu$
 in the inset to
Fig.\ \ref{Fig4}. For $\nu\leqslant 4$, the slope is
 proportional to $\nu$. So, $d v_{\text{NS}}/dM=\alpha_{\text{NS}} \nu$, 
and after integration we have,
\begin{equation}\label{Eq6}
v_{\text{NS}}=\alpha_{\text{NS}} \nu M + v_{\text{NS}}^0(\nu)
\end{equation}
where $v_{\text{NS}}^0(\nu)$ is determined by the evolutionary load of knowledge transfer. 
Linearity originates from perturbation of knowledge transfer when $\nu$ is small. 
Note that for $\nu=6$, the value used in most part of this paper, the linear relationship 
no longer holds, indicating that $\nu=6$ is large enough to break the linearity.

From our model, we make a prediction by calculating 
genetic distances between populations in different 
environments, using the genetic sequence $S'$. For each pair of environments, we calculate the 
fixation index $F_{\text{ST}}$ between them using Eq.\ 5.12 from \cite{weir1996genetic}:

\begin{equation}\label{Eq7}
F_{\text{ST}}=\frac{\sum_{i=1}^{L}\left[\frac{1}{2}\sum_{j=1}^{2}(p_{ij}-p_{ij}')^2-\frac{1}{2(2n-1)}\left(2-\sum_{j=1}^{2}(p_{ij}^2+p_{ij}'^2)\right)\right]}{\sum_{i=1}^{L}(1-\sum_{j=1}^{2}p_{ij}p_{ij}')}
\end{equation}
where $p_{i1}$ is the probability of the value of locus $i$ being $+1$, and $p_{i2}$ 
is the probability of the value of locus $i$ being $-1$ in the first environment. 
$p_{ij}'$ is the corresponding 
probability in the other environment. Here $n$ is the sample size drawn from 
the population to estimate $F_{\text{ST}}$, and in our case $n=18$, in accordance with the 
average sample size used in \cite{ramachandran2011test}.

 The east-west distance between 
environments $(x_1,y_1)$ and $(x_2,y_2)$ is $d_{\text{EW}}=|x_1-x_2|$, and the north-south distance is 
$d_{\text{NS}}=|y_1-y_2|$. We also calculated heterozygosities of the population
 of environment a, defined as

\begin{equation}
\text{het}_{a}=1-\frac{1}{L}\sum_{i=1}^L\sum_{j=1}^2 p_{ij}^2
\end{equation}
where $p_{i1}$ and $p_{i2}$ have the same meanings as those in Eq.\ \ref{Eq7}. 
Each fixation index $F_{\text{ST}}$ 
was regressed onto the sum of mean heterozygosity and geographic distance, which 
can be either east-west distance or north-south distance. The $R^2$ of the regression 
is around $0.9$. For each pair of environments 
$a$ and $b$, we express the $F_{\text{ST}}$ as,

\begin{eqnarray}
F_{\text{ST}}&=&c_{\text{EW}} d_{\text{EW}}+c_1\frac{\text{het}_a+\text{het}_b}{2}+c_0\\
F_{\text{ST}}&=&c_{\text{NS}} d_{\text{NS}}+c_1'\frac{\text{het}_a+\text{het}_b}{2}+c_0'
\end{eqnarray}

\begin{figure}
\begin{center}
\includegraphics[scale=0.62]{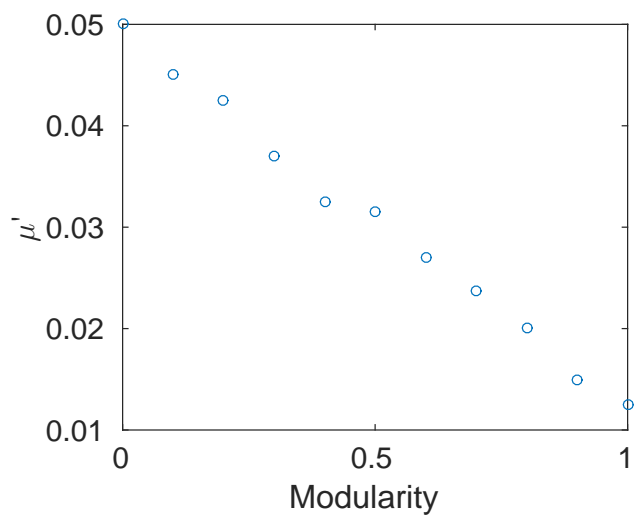}
\end{center}
\caption{\label{Fig5} Mutation rate of genetic sequence at which the $F_{\text{ST}}$ 
ratio is $1.26$ versus modularity . Other parameters are as in Fig.\ \ref{Fig1}. }
\end{figure}

 The coefficient of 
geographic distance term using east-west distance is $c_{\text{EW}}$, and $c_{\text{NS}}$ 
using north-south distance. The ratio of them, $r=c_{\text{NS}}/c_{\text{EW}}$, 
indicates the asymmetry of rate of change of genetic distance.
For humans in the Americas, the 
ratio is approximately $1.26$ \cite{ramachandran2011test}. The mutational rates of 
genetic sequences at which the $F_{\text{ST}}$ ratio is $1.26$ depend on modularity, 
as shown in Fig.\ \ref{Fig5}. The estimated mutation rate of human automosal microsatellites 
range from $10^{-4}$ to $10^{-2}$ \cite{kayser2000characteristics}. In our model, we can 
calculate the mutational rate per generation $\mu_g = \mu'\times L /2L$. So for $M=1$ the 
mutational rate is $0.005$ per
locus per generation, and for $M=0$ the mutational rate is $0.025$
per locus per generation. Thus the mutational rate for the $M=1$ case falls within the range 
of experimental results, indicating that human knowledge system is probably modular.

\section{Discussion}\label{sec:discussion}

So why is having a modular knowledge system so helpful in the human migration
process? A migrating human population must adapt knowledge quickly. 
New knowledge is generated through trial
 and error (mutation, $\mu$). Communication (knowledge corpus transfer, $\nu$)
 propagates useful new knowledge in the population. 
If the knowledge system is non-modular, however, communication 
causes confusion. This is because transfer of simply a $L/K$ segment 
 does not transfer useful information in a non-modular knowledge system. 
For example, a hunter can teach
 a wood gatherer how to hunt, including how to make stone arrowheads. If the knowledge 
system of the wood gatherer is non-modular, the hunting module can interact with 
the wood gathering module, and the wood gatherer may wrongly believe that arrow-shaped 
tools could also work for cutting trees, and replace his or her ax with arrows. For a 
modular knowledge system, this frustrating confusion will not happen, and modularity 
reduces frustration.
So if the knowledge system is modular, the population can take advantage
 of faster knowledge communication, while if the system is non-modular,
 knowledge communication can cause confusion and is deleterious
between individuals with different specializations.

For a population with a modular knowledge 
system, a smaller mutational rate of genes creates the same $F_{\text{ST}}$ ratio, 
so the evolutionary rate is higher than the non-modular counterpart when the mutational 
rates are the same. The population 
with a modular knowledge system evolves faster, and from Fisher's fundamental theorem of 
natural selection \cite{fisher1930genetical} we expect that the genetic diversity is
 higher in the more rapidly evolving population.

Why does environmental heterogeneity create an asymmetry of genetic distance in 
different directions, even if environmental change does not directly 
influence genes in our model? For a
population migrating in the north-south direction, the new environment poses severe challenges 
to the immigrants, and fewer founders may survive compared to a east-west migration. 
This founder effect 
increases the genetic distance between the immigrant population and the population 
they originate from \cite{hedrick2011genetics}. 
For the population migrating in east-west direction, much milder 
environmental changes largely reduce the founder effect, thus reducing the genetic 
distance from the original population. 

In addition to spatial heterogeneity, our stochastic model 
naturally creates temporal inhomogeneity. Even though the average fitness of a 
population changes smoothly, fitness spikes appear occasionally, corresponding 
to knowledgeable people or "heroes" in human history. Immediately after the initial 
colonization of one environment, the highest individual fitness value is more than five 
times the average fitness value of the population in our model. 
After evolution of the population for approximately $400$ generations, 
the fitness is "saturated", and the highest fitness is only 50\% better than 
the average fitness. This is consistent with our impression that more heroes emerge 
in a fast-changing society than a stagnant one.

\section{Conclusion}

In conclusion, we built a model of population migration in an asymmetric,
 two dimensional system. We have shown the vital role that modularity
 plays in the migration rates and gene flows. We have shown that a modular knowledge 
system coupled with knowledge transfer accelerates human migration. 
Our results demonstrate an east-west and
 north-south migration rate difference, and we have related environmental
 variation with longitude and latitude 
to migration rate. We have shown that the asymmetry of migration velocity 
originates from asymmetric environmental gradients. 
The asymmetry of migration velocity exists only if migration is limited by fitness. 
Predictions for asymmetry of genetic variation are in agreement 
with patterns of human gene flow in the Americas. 
Our model may be applied to other systems such as the spread of invasive species, 
cancer cells migration, and bacterial migration.

\section*{Authors' contribution}
D.W. wrote the codes and collected and analyzed the data. Both D.W. and M.W.D. 
developed and analyzed the model and drafted the manuscript. All authors gave final approval for publication.

\section*{Competing interest}
We declare we have no competing interests.

\section*{Funding}
We received no funding for this study.

\section*{Appendix}

The dynamics of evolution in one environment is described by a master equation:

\begin{eqnarray}\label{Eq8}
\frac{d P ( \{ n_{\bf a} \}; t)}{d t} &=&
\sum_{ \{ {\bf a} \} }
\bigg[
f(S_{\bf a}) (n_{\bf a}-1) \sum_{ \{ {\bf b} \ne {\bf a} \} }
 \frac{n_{\bf b}+1}{N} P(n_{\bf a}-1, n_{\bf b}+1; t)\nonumber
\\&&-f(S_{\bf a}) n_{\bf a} \sum_{ \{ {\bf b} \ne {\bf a} \} } 
\frac{n_{\bf b}}{N} P(n_{\bf a}, n_{\bf b}; t)
\bigg]\delta_{N,N^*}
\nonumber 
\\&& \nonumber +\sum_{ \{ {\bf a} \} }
\bigg[
f(S_{\bf a}) (n_{\bf a}-1)
  P(n_{\bf a}-1 ; t) -f(S_{\bf a}) n_{\bf a} 
 P(n_{\bf a}; t)
\bigg](1-\delta_{N,N^*})
\\ &&
+ \mu
\sum_{ \{ {\bf a} \} }
\sum_{ \{ {\bf b}=\partial {\bf a} \} }
\bigg[
(n_{\bf b}+1)  P(n_{\bf a}-1, n_{\bf b}+1; t) -
n_{\bf b}  P(n_{\bf a}, n_{\bf b}; t)
\bigg]
\nonumber \\ &&
+ \nu
\sum_{ \{ {\bf a} \} }
\sum_{k=1}^K
\sum_{ \{ {\bf b}, {\bf b}_k \ne {\bf a}_k  \} }
\bigg[
(n_{ {\bf a} / {\bf b}_k } +1)
 \frac{ n_ { {\bf b} / {\bf a}_k }  }{N} P(n_{\bf a}-1, 
n_{ {\bf a} / {\bf b}_k } +1; t)\nonumber
\\&&-n_{ {\bf a} / {\bf b}_k }
 \frac{ n_ { {\bf b} / {\bf a}_k }  }{N} P(n_{\bf a}, 
n_{ {\bf a} / {\bf b}_k } ; t)
\bigg]
\end{eqnarray}

Here $n_{\bf a}$ is the number of individuals with sequence $S_{\bf a}$, with
the vector index ${\bf a}$ used to label the $2^L$ sequences.
The notation $\partial {\bf a}$ means the $L$ sequences created by
a single mutation from sequence $S_{\bf a}$.
The notation ${\bf a} / {\bf b}_k$ means the sequence created by
transferring module $k$ from sequence 
$S_{\bf b}$ into sequence $S_{\bf a}$. 
Here $N^*$ is the environmental capacity of the environment.

\bibliographystyle{vancouver}
\bibliography{migration}

\end{document}